# GIARPS: the unique VIS-NIR high precision radial velocity facility in this world

Claudi R.*[a], Benatti S.[a], Carleo I.[a,b], Ghedina A.[c], Molinari E.[c], Oliva E.[d], Tozzi A.[d], Baruffolo A.[a], Cecconi M.[c], Cosentino R.[c], Fantinel D.[a], Fini L.[d], Ghinassi F.[c], Gonzalez M.[c], Gratton R.[a], Guerra J.[c], Harutyunyan A.[c], Hernandez N.[c], Iuzzolino M.[e], Lodi M.[c], Malavolta L.[b], Maldonado J.[f], Micela G.[f], Sanna N.[d], Sanjuan J.[c], Scuderi S.[g], Sozzetti A.[h], Perez Ventura H[c], Diaz Marcos H.[c], Galli A.[c], Gonzalez C.[c], Riverol L.[c], Riverol C.[c]

[a] INAF, Astronomical Observatory of Padova, vicolo Osservatorio, 5 35122 Padova Italy
[b] University of Padova, Dep. of Physics and Astronomy
[c] Fundación Galileo Galilei - INAF
[d] INAF, Astrophysical Observatory of Arcetri, Firenze Italy
[e] INAF, Astronomical Observatory of Napoli, Salita Moiariello, 16, 80131 - Napoli
[f] INAF Astrophysical Observatory of Palermo
[g] INAF, Astrophysical Observatory of Catania, via S. Sofia, 78, 95123, Catania Italy
[h] INAF, Astrophysical Observatory of Torino,

## ABSTRACT

GIARPS (GIAno & haRPS) is a project devoted to have on the same focal station of the Telescopio Nazionale Galileo (TNG) both the high resolution spectrographs HARPS-N (VIS) and GIANO (NIR) working simultaneously. This could be considered the first and unique worldwide instrument providing cross-dispersed echelle spectroscopy at a high resolution (R=115,000 in the visual and R=50,000 in the IR) and over in a wide spectral range (0.383 - 2.45 μm) in a single exposure. The science case is very broad, given the versatility of such an instrument and the large wavelength range. A number of outstanding science cases encompassing mainly extra-solar planet science starting from rocky planet search and hot Jupiters, atmosphere characterization can be considered. Furthermore both instrument can measure high precision radial velocity by means the simultaneous thorium technique (HARPS - N) and absorbing cell technique (GIANO) in a single exposure. Other science cases are also possible. Young stars and proto- planetary disks, cool stars and stellar populations, moving minor bodies in the solar system, bursting young stellar objects, cataclysmic variables and X-ray binary transients in our Galaxy, supernovae up to gamma-ray bursts in the very distant and young Universe, can take advantage of the unicity of this facility both in terms of contemporaneous wide wavelength range and high resolution spectroscopy.

## 1. INTRODUCTION

GIARPS (GIAno & haRPS-n) is a project devoted to have on the same focal station of the Telescopio Nazionale Galileo (TNG) both the high resolution spectrographs HARPS-N in the visible and GIANO in the NIR working simultaneously. This allows to have such a unique facility in the north hemisphere on duty at the TNG.

The optical coupling of both instruments is made possible by splitting the light coming from the telescope Nasmyth B focus with a dichroic. This element will be positioned in the entrance slit of the LRS (Low Resolution Spectrograph of TNG). The dichroic splits the light in two beams: the reflected visible beam feeds HARPS – N FEU (Front End Unit) while the infrared beam is injected in a preslit optical interface to GIANO. The preslit system contains all the optical elements that are necessary to bring light to the spectrograph, the slit viewer and the guiding system. In order to minimize the systematic errors and achieve Radial Velocity errors under 10 m/s an absorption cell module is also considered.

---

* *riccardo.claudi@oapd.inaf.it

## 2. SCIENCE MOTIVATION

The science case is very broad, given the versatility of such an instrument and the large wavelength range. A number of outstanding science cases encompassing mainly extra-solar planet science starting from rocky planet search and hot Jupiters, atmosphere characterization can be considered. But not only, also young stars and proto-planetary disks, cool stars and stellar populations, moving minor bodies in the solar system, bursting young stellar objects, cataclysmic variables and X-ray binary transients in our Galaxy, supernovae up to gamma-ray bursts in the very distant and young Universe, can take advantage of the unicity of this facility both in terms of contemporaneous wide wavelength range and high resolution spectroscopy.

## 3. THE VISIBLE ARM: HARPS-N

Since April 2012 the high resolution echelle spectrograph HARPS-N is part of the equipment of the Telescopio Nazionale Galileo[1]. HARPS-N is the Northern counterpart of HARPS[2], mounted at the ESO 3.6 m telescope in La Silla (Chile). Both of them allow the measurement of radial velocities with the highest accuracy now available, thanks to a precision below 1 m/s.

HARPS-N is mounted at the Nasmyth B focus of TNG, and it covers the spectral domain from 390 to 690 nm with a mean resolution of 115,000. The instrument is composed by two main parts: the spectrograph which is located in the ground floor of the telescope and the Front End and Calibration Unit (FEU) which is mounted on the telescope Nasmyth B fork. These two modules are connected through an optical fiber link. The FEU is the first part of the instrument, where the incoming light from the telescope and from the calibration unit (including the lamps for the reference source) is conditioned and collimated into the fibers. The star is maintained in the fiber thanks to the tip-tilt mirror acting together with the autoguider system. The octagonal fiber link increases the light scrambling effect and guarantees a very high precision in radial velocity measurement, since it minimize the spectrograph illumination changes due to the positioning error of the star in the fiber entrance. The two fibers have 1 arcsec aperture each: one is dedicated to the scientific object and one is used for reference (background sky or Th-Ar calibration lamp). The fiber entrance is re-imaged by the spectrograph optics onto a 4k×4k CCD, where echelle spectra of 69 orders are formed for each fiber. The spectrograph is mounted on a nickel plated stainless steel mount and contains no moving parts.

One of the peculiar characteristics of the spectrograph is its extraordinary instrumental stability, achieved thanks to the particular care taken to minimize the sources of instability. Accurate control systems avoid instrumental drifts due to variations of temperature and atmospheric pressure as well as vibrations from the floor.

To reach such precise radial velocities measurements, besides the removal of all the possible instrumental drifts, HARPS-N also guarantees an accurate localization of the wavelength in the detector with the simultaneous reference technique.

For this purpose HARPS-N is equipped with the two fibers. During scientific observations the first fiber is fed with the star light, and on this spectrum the stellar radial velocity is computed by referring to the wavelength solution determined at the beginning of the observing night. The second fiber is illuminated with the same spectral reference all the time, during wavelength calibration and scientific exposures. If an instrumental drift occurs, the simultaneous reference spectrum on the second fiber measures it.

The HARPS-N integrated pipeline[3] provides to the observer a complete reduced data set only 25 seconds after the end of the exposure. The data reduction pipeline takes into account the data images (calibration, bias, dark and scientific), performs quality control on them and executes a complete data reduction. The result is a set of data including reduced, wavelength-calibrated spectra, radial velocities, S/N etc.

## 4. THE NIR ARM: GIANO

At the beginning of 2015 TNG offered for the first time to the scientific community GIANO, its new NIR high resolution echelle spectrograph. After the commissioning and science verification observing runs in 2013 and 2014, GIANO demonstrated its capability to fulfill the required performances, reaching for instance a satisfying accuracy for the radial velocity measurements (~10m/s). A single exposure with GIANO produces a spectrum ranging from Y to K band (0.95 – 2.45 μm) with a resolution of 50,000[4].

GIANO is currently mounted at the Nasmyth-A focus of TNG and fed by IR-transmitting ZBLAN fibers. The instrument is composed by two main modules: the cryogenically cooled spectrograph and the warm preslit and interface system. The preslit system also includes a fiber mechanical agitator and rackmount with the electronics and the calibration lamps. An additional rackmount contains the detector warm electronics and controls.

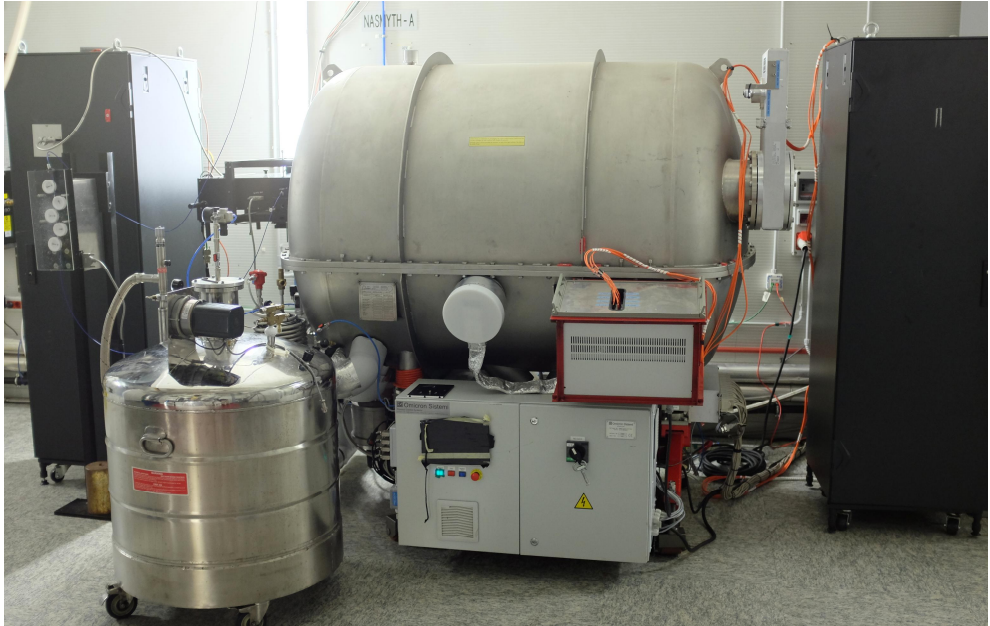

Figure 1: GIANO vacuum vessel, fiber agitator, LN2 tank and the electronics in the Nasmyth-A of TNG.

The GIANO spectrometer is mounted on a rigid aluminum bench thermally connected to a LN2 tank. Following the light path of the instrument from the entrance window, the spectrometer includes a flat window, a cold stop, a filters wheel, a slit, the spectrometer optics (7 mirrors, 3 prisms, 1 grating) and a $2k^2$ HgCdTe detector array.
All these elements are included inside a vacuum chamber (see Figure 1) which is permanently connected to all the sub-systems (pipelines, valves, pumps, sensors, PLC), necessary to create, maintain, monitor and control the vacuum and the cryogenic status of the spectrometer. All the operations are performed and supervised by the PLC, controlled through a dedicated panel. The spectrometer optical design is based on a three mirrors anastigmatic combination used in double-pass, which acts both as collimator and camera. The collimated beam is 100mm and the system focal length is 420 mm. The dispersing element is a commercial 23.2 ll/mm R2 echelle grating working at a fixed position in a quasi-Littrow configuration with an off-axis angle of  5 degrees. A system of prisms from high dispersing IR optical materials acts as cross-disperser. The optical performances of the system are excellent, with >80% en-squared energy within one pixel over most of the array area.

### 4.1 GIANO fibers:

The GIANO preslit boxes are optically connected by a couple of fibers optics, assembled in the same SMA connector and used to simultaneously measure the target and the sky. The core diameter of each fiber is 85 μm (1 arcsec on sky),the distance between the two fiber centers is 250 μm (3 arcsec on sky). A third fiber is used for calibration, and can be illuminated either by a halogen lamp for flat-field or by an U-Ne lamp for wavelength calibration. The introduction of the fibers, due to constraints imposed on the telescope interfacing during the precommissioning phase, has significantly reduced the end-to-end efficiency of the instrument. It also introduced a non repeatable spectral modulation that cannot be corrected by flat-fielding limiting thus the signal to noise ratio achievable in the spectra, regardless of the brightness of the star and integration time. For this reason GIANO uses a mechanical agitator to decrease the effect of fiber modal noise. This mechanism works quite well for diffuse sources like the calibration lamps, but for observation of scientific targets, the modal noise is amplified by effects related to the non-uniform illumination of the fiber (which also depends on the seeing conditions and on the tracking/guiding performances of the telescope). In stellar spectra acquired in non optimal observing conditions the residual modal noise can be as high as a few percent, especially at longer wavelengths, thus limiting the overall signal-to-noise ratio.

### 4.2 GIANO observations

GIANO has one observing mode at R~50,000 and can acquire spectra either of astrophysical objects and sky simultaneously, or of calibration lamps (halogen for flat-field and U-Ne for wavelength calibration) and dark frames. The echellogram (see Figure 2, left) has a fixed format and includes the orders from 32 to 81, covering the 0.95-2.45 μm wavelength range. It has a full spectral coverage up to 1.8 μm, while at the longest wavelengths the spectral coverage is about 75%. Due to the image slicer, each 2D frame contains four tracks per order (see Figure 2, right).

The read-out noise of the detector is very low: 5 e- for a single double-correlated exposure. For most applications the noise performances are dominated by the dark-current (0.05 e-/sec).

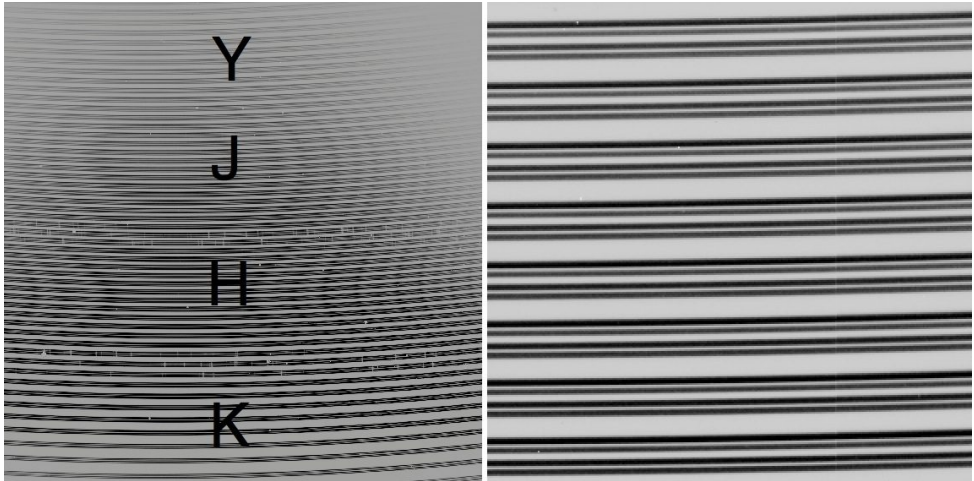

Figure 2: GIANO echellogram and photometric band position

Observations of science targets are generally performed by nodding-on-fiber, i.e. target and sky are taken in pairs and alternatively acquired on fiber A and B, respectively. From each pair of exposure an (A-B) 2D-spectrum is computed, then extracted and summed together for an optimal subtraction of the detector noise and background. The positive (on A fiber) and negative (on B fiber) spectra of the target star are shown in Figure 3.

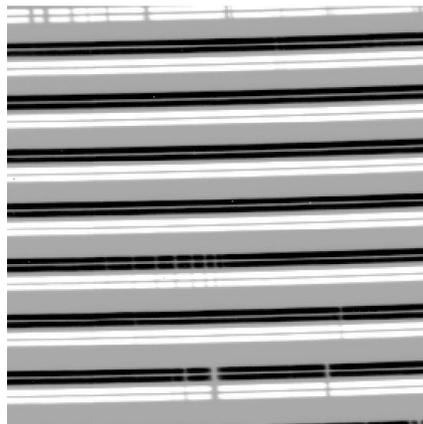

Figure 3: Science 2D-spectrum obtained with the sum of each pair of A-B spectra.

## 5. GIARPS

GIARPS (GIAno and haRPS) is a project that will allow to have simultaneous observation with both HARPS-N and GIANO. This allows to observe a wide spectral range from 0.390 to 2.45 µm at high spectral resolution: 115,000 in the visible and 50,000 in the NIR. Besides the two instrument will be also able to work alone. In summary GIARPS has three observing modes: i) HARPS-N only (maintaining the actual optical configuration with the already existing mirror); ii) GIANO only; iii) both GIANO and HARPS-N splitting the light with the dichroic (see next paragraph).
The GIARPS optics is mainly thought as the preslit optics for GIANO, it is fully described in Tozzi et al.[15]. With reference of Figure 4 a brief description is given in the following.

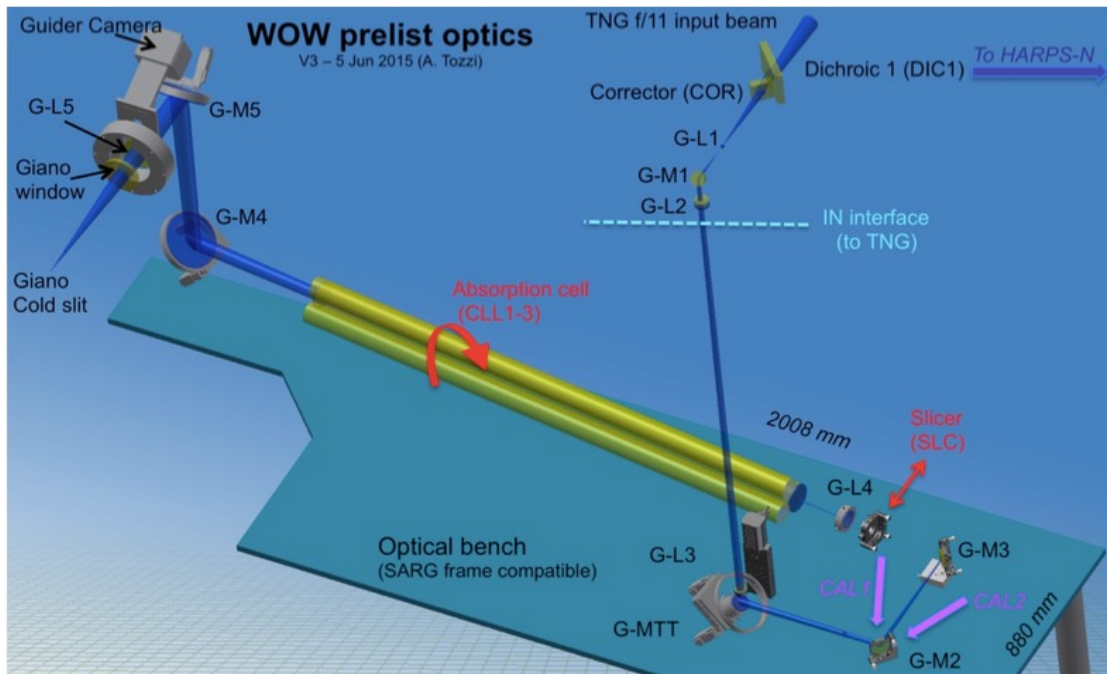

Figure 4: Scheme of the GIARPS optics necessary to fed GIANO and use it simoultaneously with HARPS – N (AD1).

The first element (DIC1 in Figure 4) is the dichroic that reflects the visible light toward HARPS-N, and transmits the IR light to GIANO. This dichroic is mounted on a slide that can select positions enabling the preferred observing mode. In the case of GIARPS observing mode the dichroic is inserted and subdivides the light in order to feed both instruments. The visible light goes toward HARPS-N Front End, while the IR light is directed towards the re-imaging module (G-L1; G-M1 and G-L2 in Figure 4) that create an intermediate focus below the de-rotator, in the volume previously allocated to SARG[5]. The light then is redirected towards the G-L3 module in order to correct and optimize the focusing of the stellar image onto the slit of the NIR spectrograph. Just after G-L3 there is a tip tilt mirror (G-MTT). The next optical elements, G-M2 (rotatory mirror) and G-M3 (fixed mirror) are used to select the calibration input from the calibration unit. The latter is beneath the preslit plane and is equipped with a U-Ne lamp, halogen lamp and a Fabry Perot selectable by a translation stage. A slicer (SLC) can be inserted in the optical path in order to enhance the efficiency during nights with poor seeing conditions. After that, mounted on a rotating stage, we use absorption cell useful for high precision radial velocities. To minimize systematic errors the gas cell should be filled with gases at the lower possible pressure, because the lines are intrinsically narrower and because the pressure-induced line-shift becomes much less important at lower pressures. This implies that, for a given mixture of gases, a long cell filled at low pressure should be always preferred to a shorter cell filled at higher pressures. For this reason we propose to use the maximum space/length available within the volume previously allocated to SARG (see Figure 4). Finally the light is brought inside the GIANO dewar by means a set of optics (the periscope G-M4/G-M5 plus the re imaging lens G-L5) that allow also to focus the light on to the entrance slit of the spectrometer. A slit viewer allows during daytime to find the exact position on the guider camera on which to center the star during observation in nighttime.

A rigid structure is foreseen to fasten the GIANO dewar to the fork of the TNG. The simple way to fix GIANO to the telescope is keeping the cryostat axis parallel to the elevation axis of the telescope (see Figure 5). All the structure will be fastened to the telescope in three points: two on the fork and one on the SARG gage (see Figure 5). The structure has been thought in order to sustain a burden of about 2000kg. The structure is enough rigid in order to not add vibration modes to those naturally generated by the movements of the telescope (Jitter, tracking etc.)[15].

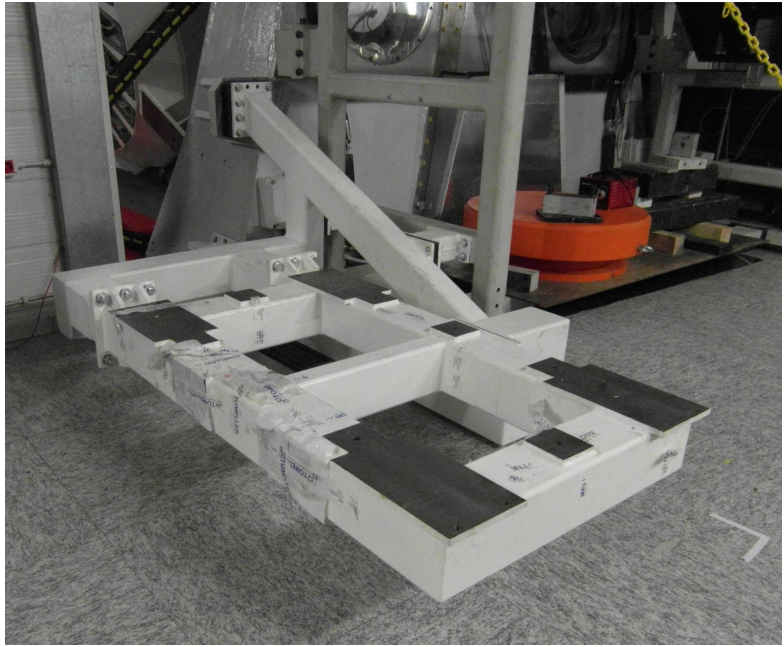

Figure 5: GIARPS optics dimension with respect to the TNG.

## 6. RADIAL VELOCITIES WITH GIARPS

The most favorable targets for radial velocity measurements are solar-type stars (F, G, and K spectral types). They are generally observed at visible wavelengths for several reasons[6]: these stars are bright at wavelength shorter than 1$\mu m$ (visible region) where the spectra are rich in deep and sharp spectral lines, so a good Doppler shift measurement is possible; spectrograph technology operating in the visible region is more advanced relative to instruments operating at other wavelengths.

In the last years less massive stars, M dwarfs, have become more interesting targets for various reasons, one of these being that M-dwarfs are more likely to host rocky planetary companions[6]. In order to find habitable planets in orbit around solar-type stars, the RV technique has to achieve a precision of 0.1 m s$^{-1}$. This constraint is released searching around less massive stars, because the reflex motion of the host stars due to the gravitational pull of the exoplanet is higher and more easily detectable than in the case of more massive stars. Moreover very cool stars such as M-dwarfs are the most numerous stars in the Galaxy[7] and these stars have closer-in habitable zones than higher-mass stars[8]. This makes finding such planets easier: the small separation and shorter periods make the amplitude of the variation of RV large and therefore the temporal stability of the instrument is less constraining. The main problem with M-dwarfs is that they are much fainter at optical wavelengths, because they have effective temperatures of 4000 K or less, and they emit most of their spectral energy at wavelengths longer than 1 $\mu m$, so they can be better observed in the near-infrared region. We know that RV signals can be induced by surface inhomogeneities, for example stellar spots[9], so a planet discovery can be confused with a variation due to such effects. An advantage of radial velocities measured from NIR spectra is that the jitter related to activity is reduced relative to visible measurements, because in the NIR the contrast between stellar spots or plagues and the rest of the stellar disk is reduced. Provided that RV can be measured with enough accuracy from NIR spectra, a comparison between variations of RV measured in the optical and NIR can establish the origin of the RV variations in an unambiguous way. For all these reasons there is a raising interest for measuring high precision RVs from NIR spectra.

The RV of a target could be measured with GIARPS on the whole range exploiting the Th simultaneous technique in the visible (the usual way of work of HARPS-N[1]) and in the NIR with GIANO exploiting the CCF method on the telluric lines or, for more precise RV measurement using an ammonia absorption cell[10].

The CCF method is performed by cross-correlating the spectrum with a mask. This is a vector with dimension equal to the observed spectrum, whose components are all zero, except those for which the condition $|\lambda_{spectrum} - \lambda_{line,i}| < step$, is satisfied, where $step = \lambda_1 - \lambda_0$, $\lambda_{spectrum}$ is the wavelength for the spectrum and $\lambda_{line,i}$ is the wavelength of the mask lines. In general the list of lines should include as many lines as possible in order to maximize the RV signal. To this purpose, we constructed two suitable digital masks that include about 2000 stellar lines, and a similar number of telluric lines[11].

RV determinations include the following steps: preparation of files including evaluation of the correction to the barycenter of the solar system; normalization of spectra; cross correlation of individual orders with the appropriate masks (both stellar and telluric spectra) with derivation of individual CCF; weighted sum of the CCFs; derivation of RVs for both stellar and telluric spectra along with the internal errors; derivation of high precision RVs; derivation of the bisector of the CCF and of the bisector velocity span (for both stellar and telluric spectra). The use of this technique allows to reach a precision of about 10 m s$^{-1}$ for bright stars (H≤5 mag) and about 70 m s$^{-1}$ (for fainter stars, typically H~9 mag).

The introduction of the absorption cell allows instead to reach a better precision of 3 ms$^{-1}$, due to the fact that absorption cell is more reliable in comparison with the instability of the telluric spectrum.

## 7. FUTURE ACTIVITIES AND CONCLUSIONS

Once GIARPS will work routinely at the telescope, TNG will have a high resolution spectroscopy station that will be unique in the northern hemisphere and up to the commissioning of NIRPS at the 3.6m ESO Telescope, the unique in this world. The flexibility of the three observing modes of GIARPS: HARPS-N alone, GIANO alone and GIARPS itself will allow users to select the best wavelength range useful for their preferred science case. From small bodies of the Solar System to the search for extrasolar planets will be the major science cases. For the latter, GIARPS will be the unique facility in this world that will allow to have simultaneously high precision radial velocity measurements in VIS (HARPS-N) and NIR (GIANO) wavelength range covering from 0.390 μm to 2.5 μm. This is very important for search for planets around active stars. In fact activity can generate a RV jitter that could be also periodic, mimicking the signal of an orbital planet. A lot of these cases happened in the previous years, one for all is the case of TW Hya[12,13]. The star is a young active star for which Setiawan et al [12] claim to find a Jupiter mass planet. Once the star was observed with CRIRES[14], the resulting RV measurements show that the amplitude of RV variations diminishes revealing the true nature of the signal[13]. A wide wavelength range on which measure RV will allow to unveil impostor signals.

The commissioning and science verification of GIARPS are foreseen at the end of 2016 and beginning of 2017, in order to, hopefully, offer the facility to the community for the AOT35 observing period at TNG.